\documentclass[preprint,12pt]{elsarticle}

 \usepackage{graphics}
 \usepackage{graphicx}

\usepackage{amssymb}

\journal{Physics Letters A}

\begin{document}

\begin{frontmatter}



\title{Dynamics of a quantum oscillator strongly and off-resonantly coupled with a two-level system}


\author{Titus Sandu}
\ead{tsandu@biodyn.ro}

\address{The International Centre of Biodynamics,
Intrarea Portocalelor Street, Nr.1B, District 6, Bucharest, Postal Code 060101, Romania}

\begin{abstract}
Beyond the rotating-wave approximation, the dynamics of a quantum oscillator interacting strongly and off-resonantly with a two-level system exhibit beatings, whose period equals the revival time of the two-level system. 
On a longer time scale, the quantum oscillator shows collapses, revivals and fractional revivals, which are encountered in oscillator observables like the mean number of oscillator quanta and in the two-level inversion population. Also the scattered oscillator field shows doublets with symmetrically displaced peaks.
\end{abstract}


\end{frontmatter}


\section{Introduction}

The two-level system
interacting with a quantum oscillator has been intensively studied both theoretically 
and experimentally. 
The model is used in a wide range of phenomena, especially in 
atomic physics where it describes a
two-level atom coupled to a quantized electromagnetic 
field \cite{Scully96,Raimond01}. 
It also describes the dynamic properties of mixed valence systems with 
an electronic transfer from one part of molecule to another \cite{Piepho78,Prassides91,Kahn88}.
Such systems are coupled vibronically and they are in degenerate or 
quasi-degenerate electronic states. The Hamiltonian describes also the ammonia molecule as a 
pseudo Jahn-Teller system \cite{Bersuker89}. In solid state physics this model 
describes many situations such as an electronic point defect in a 
semiconductor \cite{Toyozawa63} or the interaction of a dipolar impurity with the 
crystal lattice \cite{Estle68}. 

More recently the same physical model has been extended to "artificial atoms" in 
a condensed-matter environment with superconductor \cite{Chiorescu04,Wallraff04,Johansson06,Schuster07} 
and semiconductor systems \cite{Hennessy07}. These tiny solid-state devices like flux 
lines threading a superconducting loop, charges in Cooper pair boxes, and single-electron spins 
exhibit quantum-mechanical properties which can be manipulated by currents 
and voltages \cite{Vion02,Yu02,Golovach02}. Thus the vacuum Rabi splitting has been demonstrated in a Cooper pair 
box resonantly coupled to a cavity mode \cite{Wallraff04}, whereas in the dispersive regime, the photon 
quantum number has been probed \cite{Schuster07}.

The solid-state devices offer wider regimes for the coupling strength between the two-level system 
and a quantum oscillator. Typically, the coupling strength in atomic systems is 
$g \mathord{\left/ {\vphantom {g \omega }} \right. 
\kern-\nulldelimiterspace} \omega = 10^{ - 7} - 
10^{ - 6}$, where $\omega $ is the frequency of the oscillator, and $g $ is the coupling between 
the two-level system and the oscillator \cite{Raimond01}. Similar 
dipolar coupling in Cooper pair boxes and Josephson charge qubits is $3-4$ order of magnitude larger than 
the coupling in atomic systems \cite{Wallraff04,Wallraff05}. In contrast to the dipolar coupling, the 
capacitive and inductive couplings 
show even larger coupling strengths \cite{Chiorescu04,Armour02,Irish03}.

When the two-level system is conceived as a quantum bit (qubit) which interacts with a quantum oscillator 
inside a cavity, the dispersive (off-resonant) regime greatly reduces the coupling of the qubit 
with the environmental noise \cite{Blais04,Siddiqi06}. The dispersive regime can be defined as the limit where
qubit cavity detuning is larger than the coupling ($g \ll \delta$). Moreover, recently, it has been argued that the regime 
$g \mathord{\left/ {\vphantom {g \omega }} \right. \kern-\nulldelimiterspace} \omega \approx 1$ could be 
within the experimental realm \cite{Irish05}. Accordingly, the dynamical behaviour of the qubit has 
been explored for the regime 
$v < \omega$ 
($2v$ is the frequency 
of the transition between the two levels) \cite{Irish05,Sandu06}. The dynamics of the two-level 
system show for moderate coupling ($g < \omega $)  collapse and 
revival of the wave-function~\cite{Irish05}, whereas for very strong coupling ($\omega < g$)~\cite{Sandu06} the dynamical 
behaviour is a quantum version of the Landau-Zener model~\cite{Landau32,Zener32}. Collapses and revivals, the adiabatic 
approximation, and the formal connection between the two-level system interacting with a quantum oscillator and the 
Landau-Zener model has been discussed in \cite{Larson07}. The underlying assumption in deducing the connection with 
the Landau-Zener model is a large separation between the two displaced harmonic potential curves generated by 
the interaction between the two-level system and the oscillator \cite{Sandu06}. Large separation implies large $g$, 
while for moderate or small $g$ the interference between the dynamics on the two displaced harmonic potential curves 
induces collapses and revivals \cite{Irish05}.

Although the dynamics of the two-level subsystem have been studied at length in 
the regime $v,g < \omega$, less attention has been paid to the 
dynamics of the quantum oscillator subsystem. On the other hand, system measurements can be performed 
on both subsystems, but probing the oscillator is preferred as it is easily accessible via 
weak measurements \cite{Griffith06}. The present study aims at the dynamics of the 
boson field in the regime $g<v\ll \omega$ and beyond the rotating-wave approximation (RWA). The RWA 
is widely utilized in the regime $g \ll \omega$, $|\delta| \ll \omega$, and weak field because 
the anti-rotating term, which is neglected in the RWA, has a quite small contribution to the 
evolution of the system \cite{Scully96}. Despite its frequent use in quantum  optics, 
the validity of RWA is questionable for applications where strong coupling is present. For example, it is known that the 
energy spectrum of the Rabi Hamiltonian can be approximated by its RWA counterpart only for 
sufficiently small values of the coupling strength, and that the width of this range decreases as one goes higher up 
in spectrum \cite{Irish05,Feranchuk96}. On the other hand, neglecting terms of order $g^2/\delta^2$ within the same RWA, one 
obtains the shift of the oscillator frequency and the quantum counterpart of the 
ac-Stark Hamiltonian in the dispersive regime \cite{Blais04}. In the present work we will show that the 
oscillator frequency shift, which depends on the qubit state, is obtained without invoking RWA 
and without satisfying the condition $|\delta| \ll \omega$. Moreover the oscillator frequency 
shift is asymmetrical , fact that is not found in the RWA treatment. 

The regime considered here ($g< v \ll \omega$) ensures the adiabatic decoupling of the slow motion (the two-level system) from 
the fast motion (the quantum oscillator). The adiabatic approximation shows that the decoupling occurs in 
2-dimensional subspaces \cite{Irish05}. We will review the adiabatic approximation and we will recast it in a more formal 
manner that will allows us to use a perturbative expansion for the analysis of dynamics. 
On a medium time scale, while the two-level system dynamics shows true collapses and revivals, the oscillator dynamics 
shows beatings. However, on a much longer time scale 
the oscillator itself shows true collapses and revivals accompanied by fractional revivals. Fractional revival patterns are present in various 
observables of both the quantum oscillator and the two-level system. A last interesting part is the analysis of the scattered 
oscillator field that shows  a doublet 
which is characteristic to dispersive regime.

The paper is organized as follows. Section I is the introduction. Section II presents the Hamiltonian and discusses
the adiabatic approximation. 
Section III is dedicated to the analysis of the dynamics. In the last Section the conclusions are outlined.

\section{The Hamiltonian and its adiabatic approximation}
\subsection{The Hamiltonian}

We consider the interaction of an electronic two-level system with a quantum oscillator (boson field). In the basis of the two electronic states, 
$\left| 1 \right\rangle$ and $\left| 2 \right\rangle$, the Hamiltonian can be written as ($\hbar = 1)$:

\begin{equation}
\label{eq1}
H = \frac{p^2 + \omega ^2q^2}{2} - gq\sigma _z - v\sigma _x, 
\end{equation}

\noindent
where $\sigma _{z }$ and $\sigma _{x }$ are the corresponding Pauli 
matrices, $p$ and $q$ are the oscillator coordinates. The essential parameters are 
\textit{$\omega $}, $g$, and $v$ associated with the frequency of the 
oscillator, the electron-oscillator coupling strength, and the separation of the
electron levels, respectively. 
This Hamiltonian is isomorphic with the simple pseudo 
Jahn-Teller system~\cite{Bersuker89} or the Rabi Hamiltonian~\cite{Rabi37}
[a rotation in $\sigma $ space with $e^{ - i\frac{\pi 
}{4}\sigma _y }$ or $H$ written in the basis $\frac{1}{\sqrt 2 }\left( {\left| 
1 \right\rangle \pm \left| 2 \right\rangle } \right)$]:

\begin{equation}
\label{eq2}
H = \frac{p^2 + \omega ^2q^2}{2} - gq\sigma _x - v\sigma _z.
\end{equation}

\noindent
In this form the parameter $v$ directly shows its role as the splitting of the two 
levels. A simplification of (\ref{eq2}) is the famous Jaynes-Cummings model that has 
played a very important role in understanding the interaction between 
radiation and matter in the resonant regime \cite{Jaynes63}. 
With three essential parameters, in the dispersive regime, the Hamiltonian (\ref{eq1}) exhibits a wide variety of phenomena like 
the squeezing of the boson field~\cite{Sandu03}, the collapse and revival of the wave packet~\cite{Irish05}, and the 
full quantum Landau-Zener effect \cite{Sandu06}. The simplicity of the Hamiltonian (\ref{eq1}) led to the conjecture that it has an exact solution in terms of 
known functions \cite{Reik86,Reik87} but an analytical solution is lacking.  Therefore 
approximations 
are needed to understand the dynamical behaviour of (\ref{eq1}). Throughout the rest of this paper we are going to work 
with a dimensionless time 
$\omega \,t \to t$ by setting $\omega = 1$ without the loss of generality. 
In the next subsections we will present a stationary adiabatic 
approximation of the two-level 
system strongly coupled with a quantum oscillator. 
It will be based on the displaced oscillator basis provided by the strong coupling and 
on the relationship between 
the other two parameters, oscillator frequency that is taken to be 1 and the frequency associated with the 
energy separation of the two levels, namely $v \ll 1$.

\subsection{Adiabatic approximation to the relative strong coupling $(0 \ll g < v \ll 1)$}

The Hamiltonian (\ref{eq1}) is a two-site realization of the small polaron model \cite{Holstein59}. 
In the limit $v = 0$ it can be solved exactly by the Lang-Firsov transformation \cite{LangFirsov63}: $U = e^{ - ig{\kern 1pt} p \sigma _z}$.
This unitary transformation expresses the Hamiltonian into the basis generated by the displaced oscillators 
that diagonalizes the sector $v = 0$ of the Hamiltonian (\ref{eq1}). The new Hamiltonian is

\begin{equation}
\label{eq14}
\tilde {H} = 
\frac{1}{2}(p^2 + q^2) - \frac{g^2}{2} - v\left( {{\begin{array}{*{20}c}
 0 \hfill & {e^{2 i g p}} \hfill \\
 {e^{ - 2 i g p}} \hfill & 0 \hfill \\
\end{array} }} \right).
\end{equation}

\noindent
Perturbation calculations can be 
performed as long as $v \ll 1$ \cite{Schweber67}. The net effect of 
$e^{ \pm 2 i g p}$
on the wave-function is to displace it by the amount $ \pm 2 g $. 
Thus, the effective splitting given by the third term of the Hamiltonian (\ref{eq14}) 
will be quenched, such that 
perturbation calculations on the Hamiltonian (\ref{eq14}) can be extended to larger values 
of $v$ \cite{Sandu06}. 
Going further, we cast the Hamiltonian (\ref{eq14}) in the basis 
$\frac{1}{\sqrt 2 }\left( {\left| 1 \right\rangle \pm \left| 2 \right\rangle } \right)$, 
\begin{equation}
\label{eq14adiab}
\tilde {H'} = 
\frac{(p^2 + q^2-g^2)}{2} - \frac{v}{2} \left( {{\begin{array}{*{20}c}
 {e^{2 i g p} + e^{ - 2 i g p}} \hfill & {e^{ - 2 i g p} - e^{2 i g p}} \hfill \\
 {e^{2 i g p} - e^{ - 2 i g p}} \hfill & {-e^{2 i g p} - e^{ - 2 i g p}} \hfill \\
\end{array} }} \right).
\end{equation}

\noindent
So far no approximation has been made. The perturbation calculations that consider only the diagonal part of the last term in 
Eq.(\ref{eq14adiab}) and the 
Fock states of the oscillator lead to the adiabatic approximation due to the fact 
that the perturbation which is proportional 
to $v$ is much smaller than the energy quanta of the oscillator ($v \ll 1$). We denote the states of the adiabatic 
approximation as $\{\left|{1;n} \right\rangle, \left|{2;n} \right\rangle \}$, where $1$ and $2$ are the electronic states and 
$n$ is the oscillator quantum number. It is easy to check that the off-diagonal part of the last term 
in (\ref{eq14adiab}) 
brings no contribution in the first order approximation. Furthermore, expanding the exponentials up to the second order in 
$g$, we obtain a shift in oscillator frequency. The shift depends on the state of the two-level system and it is obtained 
without using the RWA as it has been reported in Ref.~\cite{Blais04}. Moreover, in the next section we will show that,
in contrast to \cite{Blais04}, the oscillator frequency shift is uneven. In addition, the Hamiltonian (\ref{eq14adiab}) 
allows an easy perturbative approach that we will use it in the next section. Also, Eq.~(\ref{eq14adiab}) provides a more  transparent 
way to improve the adiabatic approximation and to construct a generalized rotating-wave approximation (GRWA) as the one 
that has been recently published \cite{Irish07}. In the formulation given by the Hamiltonian (\ref{eq14adiab}), the GRWA is restricted to coupling the
states $\{\left|{2;n} \right\rangle$ to $\left|{1;n+1} \right\rangle \}$, thus the eigenvalues and the eigenvectors of GRWA can be calculated 
straightforwardly.

The same adiabatic result is obtained 
if the matrix elements of Hamiltonian  (\ref{eq1}) 
are calculated directly in the basis $\left| {N_\pm } \right\rangle$ 
(Fock states) of the oscillators displaced by $\pm g$, where $N_\pm$ is the number of 
quanta of the 
state $\left| {N_\pm } \right\rangle$. The adiabatic approximation as described above amounts to 
coupling the states $\left|{N_{+}} \right\rangle$ with the states $\left|{N_{-}} \right\rangle$ 
provided that $N_{-} = N_{+}$ \cite{Irish05}. The Hilbert space of the adiabatic 
Hamiltonian is split 
in a direct sum of 2-dimensional sub-spaces 
$\{\left|{N_{-}} \right\rangle, \left|{N_{+}} \right\rangle \}$. Each sub-space 
$\{\left|{N_{-}} \right\rangle, \left|{N_{+}} \right\rangle \}$ has the following eigenstates and eigenvalues

\begin{equation}
\label{adiabwf}
 \left| {\pm ,N} \right\rangle = \frac{1}{\sqrt 2 }\left( {\left| {N_ + } 
\right\rangle \pm \left| {N_ - } \right\rangle } \right),  
\end{equation}

\begin{equation}
\label{adiabenerg}
E_{\pm ,N} = \pm v\left\langle {N_ - } \mathrel{\left| {\vphantom {{N_ - } 
{N_ + }}} \right. \kern-\nulldelimiterspace} {N_ + } \right\rangle + N + \frac{1-g^2}{2}.
\end{equation}

\noindent
The energy eigenvalues given by Eq.~(\ref{adiabenerg}) provide a very good approximation to the 
full Hamiltonian (\ref{eq14}) \cite{Irish05}. In the next section we will use the adiabatic 
approximation to explore the dynamics of the quantum oscillator and of the two-level system on a time scale of up to $O(\frac{1} {vg^4})$.

\begin{figure}
\includegraphics{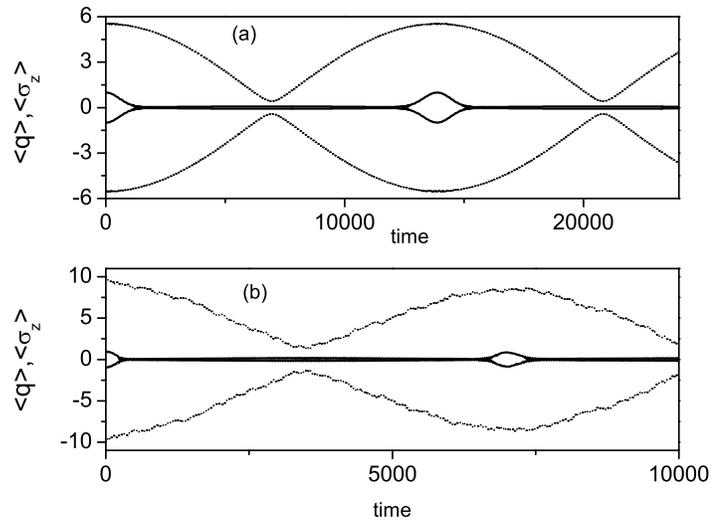}
\caption{\label{fig:2} Collapse and revivals of two-level (full line) and quantum oscillator (dotted line) systems. Only the envelopes 
are shown. There are two cases: (a) $|\alpha|^2 = 30$, $g=0.0333$, $v=0.1$;
(b)$|\alpha|^2 = 100$, $g=0.05$, $v=0.1$. }
\end{figure}

\section{Results and discussions}

The integration of the full dynamics was made with a split-operator method \cite{Sandu03}. For numerical 
applications we consider as initial conditions a wave packet of the form 
$\left| 1 \right\rangle \left| \alpha \right\rangle $, where $\left| \alpha \right\rangle $ is the coherent state \cite{Scully96,Glauber63} 
which gives the average quantum number of the bosonic part $\langle n \rangle = |\alpha|^2$. 
Collapses and revivals of the two-level wave-function  \cite{Eberly80} as well as of the oscillator wave-function are shown in Fig.~\ref{fig:2}. 
In fact, collapses and revivals of the oscillator wave-function are beatings.

\begin{figure}
\includegraphics{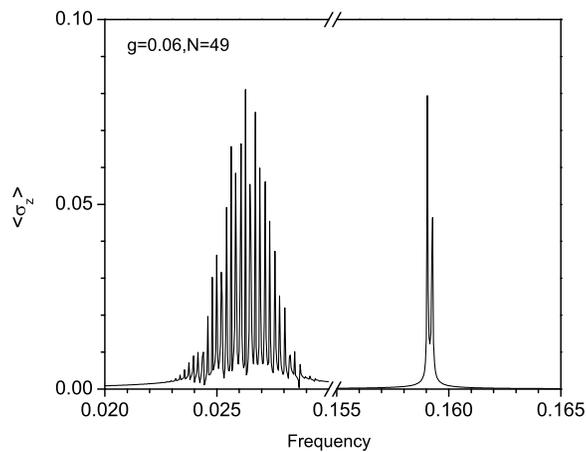}
\caption{\label{fig:Rabi} Fourier spectrum of  $\langle \sigma _z \rangle$. Low frequency components are characteristic to the 
Fock states of the  oscillator, while the high 
frequency feature shows the coupling to the quantum oscillator.}
\end{figure}

The collapse and revival 
of the wave-function in the two-level system can be 
understood in terms of the composite Rabi frequency $v_{Rabi}$, i.e. the total Rabi frequency is made from contributions of 
many different Rabi frequencies of individual eigenstates of the quantum oscillator (Fig.~\ref{fig:Rabi}). When enough time has 
passed, different oscillatory terms get out-of-phase and cancel out each other. This is the quantum collapse and it is 
related to the spread of the Rabi frequency distribution that is provided by the wave packet. The revival or the restoration of the two-level wave 
function to nearly its initial value is due to the discrete nature of the Rabi frequency distribution and the 
revival time is related to the difference of the neighbor Rabi frequencies. Revivals are purely quantum phenomena, 
arising from the fact that the quantum number distribution of the oscillator is not continuous, whereas the 
collapse is completely classical. The collapse and revival of the two-level system is also shown by the adiabatic solution 
given by Eqs.~(\ref{adiabwf}) and (\ref{adiabenerg}), whose dynamics preserves the general features of the 
collapse and revival phenomenon as just sketched above \cite{Irish05}. The collapse and revival times can be evaluated by expanding the 
off-diagonal exponentials in Eq.~(\ref{eq14}). Thus retaining the terms up to second order in $p$ and sandwiching the Hamiltonian 
(\ref{eq14}) between the adiabatic states $\{\left|{1;n} \right\rangle, \left|{2;n} \right\rangle \}$ we can estimate the 
variation of the Rabi frequency from level $n$ to level $n+1$ 

\begin{equation}
\label{varRabi}
 v_{Rabi}(n+1) - v_{Rabi}(n) = 4vg^2. 
\end{equation}

\noindent
Equation (\ref{varRabi}) together with the spread of the wave packet $\alpha = \sqrt N$ provide the revival time 

\begin{equation}
\label{revival-time}
 t_{1rev} = \frac{\pi }{2vg^2} 
\end{equation}
and the collapse time

\begin{equation}
\label{collapse-time}
t_{1coll} = \frac{\pi }{4{\kern 1pt} \alpha \,v{\kern 1pt} g^2}. 
\end{equation}

\begin{figure}
\includegraphics{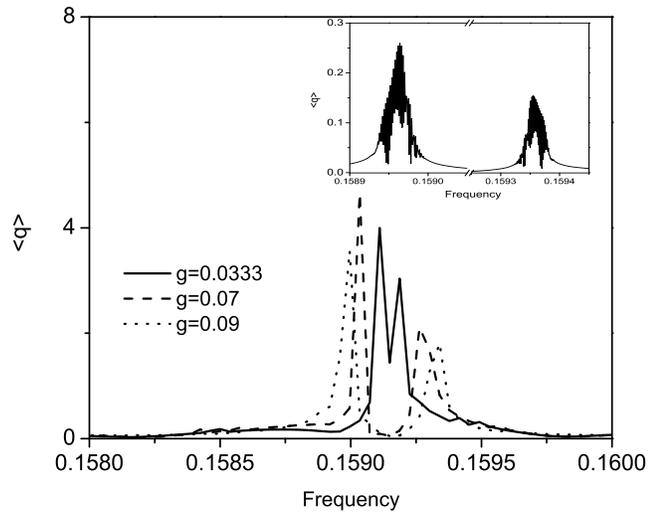}
\caption{\label{fig:3} Spectra of the oscillator coordinate $\langle q \rangle$ around the unperturbed oscillator frequency for various coupling strengths $g$ and $v=0.1$. 
They show splittings due to the interaction with the two-level system. 
The finer resolution that is presented in the inset indicates the splitting is dependent on the Fock states of the oscillator.}
\end{figure}

The dynamics of quantum oscillator, however, show beatings. 
The beatings of the quantum oscillator can be understood by analyzing the Fourier transform of oscillator coordinate. 
The spectral features of the oscillator coordinate 
$\langle q \rangle$ are plotted in Fig.~\ref{fig:3}. 
A split-like feature is exhibited in frequency by the exact dynamics of the quantum oscillator, i.e. 
the two-level system "pulls" the 
oscillator frequency that now depends on the state of the two-level system. 

Quantitative analysis of the splitting 
can be done by considering the diagonal part of the Hamiltonian (\ref{eq14adiab}) and discarding the constant term $-g^2/2$  

\begin{equation}
\label{eqadiab-1}
\tilde {H'}_{adiab} = 
\frac{(p^2 + q^2)}{2} - \frac{v}{2} \left( {{\begin{array}{*{20}c}
 {e^{2 i g p} + e^{ - 2 i g p}} \hfill & {0} \hfill \\
 {0} \hfill & {-e^{2 i g p} - e^{ - 2 i g p}} \hfill \\
\end{array} }} \right).
\end{equation}

\noindent
The expansion of the exponentials in Eq.~(\ref{eqadiab-1}) up to the fourth order 
in $p$ provides the approximate Hamiltonian 

\begin{equation}
\label{eqadiab-1_approx}
\tilde {H}' = \left({{\begin{array}{*{20}c}
 {\frac{1}{\Omega _ + ^2 }\frac{1}{2}(p^2 + \Omega _ + ^2 q^2) - v - 
\frac{2}{3}vg^4p^4} \hfill & 0 \hfill \\
 0 \hfill & {\frac{1}{\Omega _ - ^2 }\frac{1}{2}(p^2 + \Omega _ - ^2 q^2) + 
v + \frac{2}{3}vg^4p^4} \hfill \\
\end{array} }}\right) + O\left({p^6}\right),
\end{equation}

\noindent 
with $\Omega _\pm = \frac{1}{\sqrt {1\pm 4vg^2} } \approx 1 \mp 2vg^2$. The validity 
of the expansion (\ref{eqadiab-1_approx}) is bounded to maximal quantum number $n$ given 
by $vn^2g^4<1$, which sets the range of validity 
beyond any practical application as long as $g<v \ll 1$. For instance, 
a typical example $v=0.1, g=0.1$ (largest value of $g$ when $v$ is set) implies $n<316$, 
which is more than enough for a practical application. 

Thus the splitting calculated 
to the second order in $g$ is $\Omega _ - -\Omega _ + =4vg^2$, i.e. it is just the variation of the Rabi frequency from level 
$n+1$ to  level $n$ that is given by Eq. (\ref{varRabi}). Therefore the minimum amplitude 
of the oscillator coordinate due to the beatings occurs at the half of the revival time of 
the two-level wave-function, hence 
the oscillator amplitude may indicate that the two-level system is in a pure state \cite{Gea-Banacloche90}. 
In addition, the fractions $\frac{1}{\Omega _ {\pm} ^2 }$ 
in Eq. (\ref{eqadiab-1_approx}) show the fact 
that the splitting is uneven. This fact is also found in the full numerical calculations shown in Fig.~\ref{fig:3}. 
We note here that the asymmetrical splitting is not found with a RWA treatment \cite{Blais04}. 

Spreading in quantum number $n$ induces broadening in frequency pulling, thus the splitting is 
well separated only for quite large $g$ ($\approx 0.1$). In the inset of Fig.~\ref{fig:3} it is presented a finer resolution plot of splitting 
which shows its dependence on the Fock states of the oscillator. The finer resolution will lead us to longer time scale 
dynamics and to true collapses and revivals of the oscillator wave-function. 
The same beatings of the 
oscillator coordinate can be also seen in another regime, $v > 1$ for a spin coupled to a nanomechanical resonator \cite{Xue07}. In that regime,  
$v > 1$, the adiabatic motion of the oscillator undergoes a frequency change from 
$1$ to $\sqrt{1 - g^2/v}$ \cite{Sandu03}, therefore a classical picture of beatings is established with the driving force at 
the frequency of the unperturbed oscillator and the oscillating system at the frequency $\sqrt{1 - g^2/v}$. 

The analysis can be carried out further by considering the approximate wave-function 
of the Hamiltonian (\ref{eqadiab-1_approx}),

\begin{equation}
\label{approx_wave-function}
\left( {{\begin{array}{*{20}c}
 {\left| {\psi _1 } \right\rangle } \hfill \\
 {\left| {\psi _2 } \right\rangle } \hfill \\
\end{array} }} \right) \approx \left( {{\begin{array}{*{20}c}
 {e^{-i\varphi _1 t}\sum\limits_n {C_n \;e^{-i\,n\,\left( {1 + 2vg^2 - vg^4} 
\right)\;t}e^{-i\,n^2\,\left( { - vg^4} \right)\;t}\left| n \right\rangle } } 
\hfill \\
 {e^{-i\varphi _2 t}\sum\limits_n {C_n \;e^{-i\,n\,\left( {1 - 2vg^2 + vg^4} 
\right)\;t}e^{-i\,n^2\,\left( {vg^4} \right)\;t}\left| n \right\rangle } } 
\hfill \\
\end{array} }} \right),
\end{equation}

\noindent
where $C_n$ are the coefficients of the initial wave packet, $\phi_1=-v+ \frac{1} {2} (1+2vg^2-vg^4)$, and 
$\phi_2=v+ \frac{1} {2} (1-2vg^2+vg^4)$. The validity of the wave-function (\ref{approx_wave-function}) is controlled 
by the stronger condition $v n^2 g^4 t <1$ because the energy corrections are included in the phases of the fast oscillating terms. 
The approximations used in \cite{Feranchuk96} and \cite{Irish07} would describe better the dynamics on such large 
time scales and on wider range of parameters. When using those better approximations, however, there are no simple relationships 
between the time scales of the dynamics and the parameters of the system. Thus one has to invoke approximations like (\ref{eqadiab-1_approx}) and 
(\ref{approx_wave-function}) in order to have meaningful relationships between the time scales of the dynamics and the parameters of the system. 
It is worth mentioning that a weak coupling limit is used in Ref. \cite{Irish05} for the evaluation of the revival time. The wave-function 
(\ref{approx_wave-function}) resembles the wave-function of an anharmonic oscillator 
and it suggests the analysis of longer time scale 
dynamics \cite{Robinett04}. The approach of studying collapses and revivals as outlined in \cite{Robinett04} 
is, however, suitable for single potential surfaces. Once the potentials are coupled, like in the present model, 
the analysis is more elaborate \cite{Wang08}. Thus we pursue an explicit evaluation of the expectation
 values of several observables like  

\begin{equation}
\label{expect_sigma}
\left\langle {\sigma _z } \right\rangle \approx 2 Re\left( \left\langle {\psi _2 } 
\mathrel{\left| {\vphantom {{\psi _2 } {\psi _1 }}} \right. 
\kern-\nulldelimiterspace} {\psi _1 } \right\rangle \right) = 2 Re\left( e^{-i\left( {\varphi _1 
- \varphi _2 } \right)t}\sum\limits_n {\left| {C_n } \right|^2e^{-in\left( 
{4vg^2 - 2vg^4} \right)t}e^{ 2in^2vg^4t}} \right), 
\end{equation}

\noindent
while the expectation value of $q$ up to a residual term $g \langle \sigma_z \rangle$ has the following form 

\begin{equation}
\label{expect_q}
\begin{array}{l}
 \left\langle q \right\rangle - g\left\langle {\sigma { }_z} \right\rangle \approx \left\langle {\psi _1 } \right|q\left| 
{\psi _1 } \right\rangle + \left\langle {\psi _2 } \right|q\left| {\psi _2 } 
\right\rangle =  \\ 
 \frac{1}{\sqrt 2 }\sum\limits_n {\sqrt n C_{n - 1}^\ast C_n e^{-i\left( {1 + 
2vg^2} \right)t}e^{ 2invg^4t} + \sqrt {n + 1} C_{n + 1}^\ast C_n e^{ i\left( {1 + 2vg^2 - 2vg^4} \right)t}e^{-2i\,nvg^4t}} + \\ 
 \frac{1}{\sqrt 2 }\sum\limits_n {\sqrt n C_{n - 1}^\ast C_n e^{-i\left( {1 - 
2vg^2 } \right)t}e^{-2invg^4t} + \sqrt {n + 1} C_{n + 1}^\ast C_n e^{ i\left( {1 - 2vg^2+ 2vg^4} \right)t}e^{2i\,nvg^4t}}. \\ 
 \end{array}
\end{equation}

\begin{figure}
\includegraphics{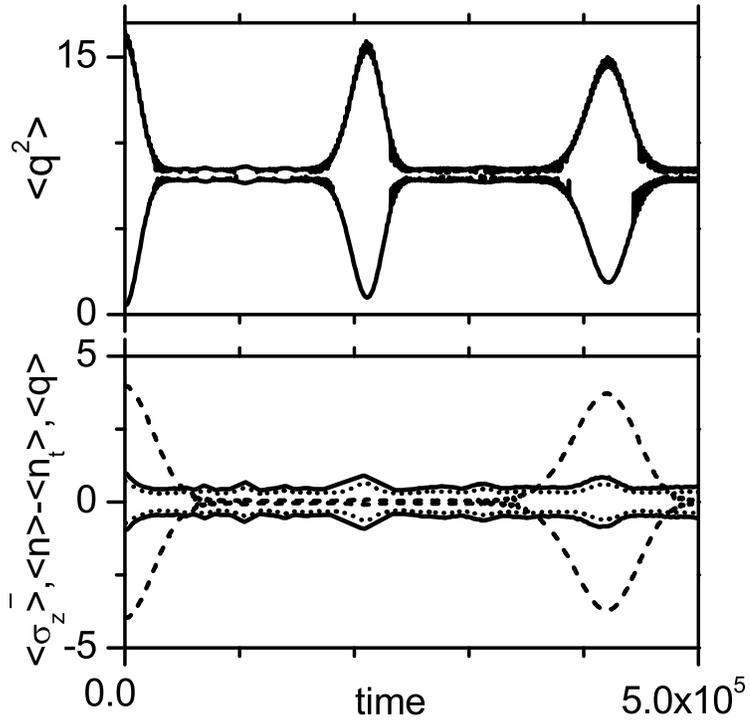}
\caption{\label{fig:22} The collapse and revival of the oscillator wave-function. The top panel depicts the time evolution 
of $\langle q^2 \rangle$ that has the revival time halved with respect to $\langle q \rangle$. The bottom panel shows the 
dynamics of $\langle q \rangle$ (dashed line), $\langle \sigma_z \rangle$ (solid line) and $\langle n \rangle -\langle n \rangle_t $ 
(dotted line). The parameters are: $|\alpha|^2 = 16$, $g=0.09$, $v=0.1$}
\end{figure}

\noindent 
Moreover, the difference between the expectation value of the oscillator quantum number $\langle n \rangle$ 
and its time average is

\begin{equation}
\label{expect_n}
\begin{array}{l}
 \left\langle n \right\rangle - \left\langle n \right\rangle _t \approx 
 2Re \left(g\left\langle {\psi _2 } \right|q\left| {\psi _1 } \right\rangle \right) = 
g\sqrt 2 Re( e^{-i\left( {\varphi _1 - \varphi _2 } \right)\,t}\times \\ 
 \sum\limits_n {e^{-4invg^2t}e^{2in^2vg^4t}\left( {\sqrt n C_{n - 1}^\ast C_n e^{-i\left( {1 - 2vg^2} \right)t} 
+ \sqrt {n + 1} C_{n + 1}^\ast C_n e^{4invg^4t}e^{ i\left( {1 - 2vg^2 + 2vg^4} \right)t}} \right)).} \\ 
 \end{array}
\end{equation}

\noindent
Here $\langle n \rangle _t$ is the time average of $\langle n \rangle$. For a time scale of order 
$O(\frac {2\pi}{4vg^2})$ the terms in Eq. (\ref{expect_sigma}) rephase, thus one recovers the revival time of 
 $\langle \sigma_z \rangle$ that is given by Eq. (\ref{revival-time}). The examination of Eqs.(\ref{expect_sigma}), 
(\ref{expect_q}), and (\ref{expect_n}) reveals a longer time scale of order $O(\frac {2\pi}{2vg^4})$ when their terms  
rephase again leading to the revival of the oscillator wave-function. In Fig.~\ref{fig:22} we have plotted the time evolution of $\langle q^2  \rangle$ 
in the top panel and of $\langle q \rangle$, $\langle \sigma_z \rangle$, and  $\langle n \rangle - \langle n \rangle _t$ in the 
lower panel. The new revival time  

\begin{equation}
\label{revival-time-oscill}
 t_{2rev} = \frac{\pi }{vg^4}, 
\end{equation}

\noindent
is the revival time of $q$. The new time $t_{2rev}$ that is determined by the nearest neighbor level spacing is twice the revival 
time of $\langle  q^2 \rangle$,
which is set by the next to nearest neighbor level spacing. Collapses and revivals on a much longer time scale have been also predicted in a  driven 
Jaynes-Cummings model \cite{Chough96}, but the observable subjected to collapses and revivals is $\langle n \rangle$ instead of $\langle q \rangle$. 
The case 
analyzed in the present work shows that, in terms of collapses and revivals, $\langle n \rangle$ behaves similarly to $\langle \sigma_z \rangle$. 
Moreover $t_{2rev}$ is similar to the fractional revival time $T_{FR}$ 
identified in the Jaynes-Cummings model \cite{Averbukh92}. 
We notice that the time scales defined by the adiabatic approximation are different from the time scale defined by RWA and Jaynes-Cummings model. 
In the Jaynes-Cummings model $t_{1rev}$ and $t_{2rev}$ depend strongly on the initial wave-packet by the mean number of oscillator quanta~\cite{Averbukh92}.

\begin{figure}
\includegraphics{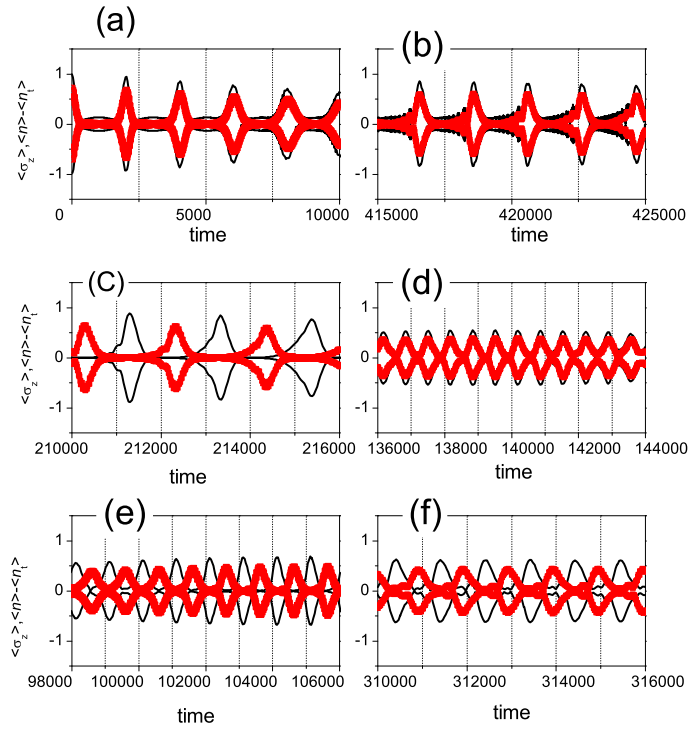}
\caption{\label{fig:33} (Color online) Fractional revivals that can be observed in 
$\langle \sigma_z \rangle$ and  $\langle n \rangle- \langle n  \rangle_t$. 
 (a) and (b) at the beginning and the end of the interval $[0, t_{2rev}]$, respectively; (c) around  $t_{2rev}/2$; (d) 
around  $t_{2rev}/3$; (e)  and (f) in 
the vicinity of $t_{2rev}/4 $ and  $3 t_{2rev}/4$, respectively. The thin black lines denote 
the dynamics of $\langle \sigma_z \rangle$ and the thick red lines  depict the dynamics of 
 $\langle n \rangle- \langle n \rangle_t$. The parameters are the same as the ones used in Fig.~\ref{fig:22}.}
\end{figure}

A close look in the interval $[0, t_{2rev}]$ exposes rich structures 
of fractional  revivals in the 
vecinity of $(p/q)t_{2rev}$ ($p$ and $q$ are mutually prime integers) \cite{Averbukh89}.
The analysis of the long time behaviour indicates us that the identification of fractional revivals in an oscillator strongly coupled 
to a two-level system can be made by simply analyzing observables like $\langle \sigma_z \rangle$ or  $\langle n \rangle$ instead 
of calculating  the information entropy \cite{Romera07}. 
Thus in Fig.~\ref{fig:33} one can see that at the beginning and the end of the interval $[0, t_{2rev}]$  the collapses and revivals of 
$\langle \sigma_z \rangle$ and $\langle n \rangle- \langle n \rangle _t$ are similar or ``in phase'', i.e. they occur at the same time. However, 
in the vicinity of $t_{2rev}/2$ the same collapses and revivals are ``out of phase'': the amplitude of $\langle \sigma_z \rangle$ is 
collapsed and the amplitude of $\langle n \rangle- \langle n \rangle _t$ is revived, or conversely the amplitude 
of $\langle n \rangle- \langle  n \rangle _t$ is collapsed and the amplitude of $\langle \sigma_z \rangle$ is revived. 
The ``in phase'' collapses and revivals occur, 
for example, in the vecinity  of $t_{2rev}/3$, 
but the revival time is one third of the revival time  (\ref{revival-time}). The above behaviour is explined by invoking  Eqs. (\ref{expect_sigma})  and 
(\ref{expect_n}). Thus at $t \approx t_{2rev}/2$ the phases brought in by the terms $e^{2invg^4t}e^{ 2in^2vg^4t}$ in Eq. (\ref{expect_sigma}) are  $0$. 
In turn the phases brought in by the terms 
$e^{2in^2vg^4t}$ in Eq. (\ref{expect_n}) are $\pi$. Consequently the collapses and revivals of $\langle \sigma_z \rangle$ are  ``out of phase'' 
with respect to collapses and revivals of $\langle n \rangle- \langle n \rangle _t$. Similar behaviour is shown 
at  $t \approx t_{2rev}/4$ and $t  \approx 3t_{2rev}/4$, while 
at $t \approx t_{2rev}/3$ the revival time of $\langle \sigma_z \rangle$ and $\langle n \rangle- \langle n \rangle _t$ is  $t_{1rev}/3$ \cite{Averbukh89}. 

\begin{figure}
\includegraphics{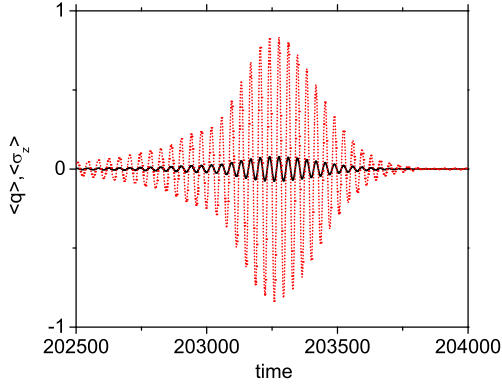}
\caption{\label{fig:44} (Color online) Full dynamics of  $\langle \sigma_z \rangle$ and  $\langle q \rangle$ in the vicinity of $t_{2rev}/2$. 
The full line depicts the dynamics of $\langle q \rangle$ and the dotted line denotes the dynamics of $\langle \sigma_z \rangle$.}
\end{figure}

Another interesting aspect is the 
frequency of $\langle q \rangle$ within collapse region as it can be seen in  Fig.~\ref{fig:44}: the frequency 
of $\langle q \rangle$ equals the  frequency of the two-level system. Meanwhile in the revival regions the frequency 
of $\langle q \rangle$ is close to the frequency of the unperturbed oscillator. The inspection of Eq. (\ref{expect_q}) 
tells us that, in the collapse region, $\langle q \rangle$ is mainly the term $g\langle \sigma_z \rangle$,  hence the 
dynamics of $\langle q \rangle$ and  $\langle \sigma_z \rangle$ are similar in the 
collapse region of the oscillator.

The approximate wave-function (\ref{approx_wave-function}) allows us to calculate also the scattered field of the oscillator by the two-level  system. 
We define the scattered field as the Fourier transform of the first order correlation function:

\begin{equation}
\label{eq_fluoresc}
I_{scatt} \propto Re\left( {\int\limits_0^\infty {dt} \int\limits_0^\infty {dt'} 
\left\langle {a^{\dag}  \left( t \right)\sigma _z \left( t \right)a\left( {t'} 
\right)\sigma _z \left( {t'} \right)} \right\rangle e^{ - i\omega \left( {t 
- t'} \right)}\theta \left( {t - t'} \right)} \right).
\end{equation}

\noindent
Direct calculations lead us to

\begin{equation}
\label{eq_fluoresc_formula}
I_{scatt} \propto \sum\limits_{n,\pm} {n\left| {C_n } \right|^2 {\delta \left( {\omega - 1 \mp 
2v \mp vg^4 \pm 4nvg^2 \mp 2n^2vg^4} \right) } }.
\end{equation}

\noindent
Unlike the spectrum of $\langle q \rangle$, the spectrum of the scattered field shows two lobs symmetrically displaced by the low frequency components of Rabi frequencies with respect to the frequency of 
the  unperturbed oscillator. The spectrum of scattered field is the counterpart to the fluorescence spectrum obtained 
by strong field pumping of a two-level system near the  resonance \cite{Scully96}. 
Thus, experimentally the scattered field can be probed in the same way like the probing of fluorescence spectrum \cite{Scully96}. 
Fluorescence spectrum has three peaks, 
one dominant central peak at the frequency of the pump field and two side peaks  symmetrically displaced 
by Rabi frequency from the central peak \cite{Mollow69}.  The spectrum given by Eq. (\ref{eq_fluoresc_formula}), however, has two lobs  
that are characteristic to a dispersive regime. Moreover, the scattered 
spectrum shows also information about the occupation of oscillator Fock states.

\section{Conclusions}
We have studied the dynamics of a quantum oscillator interacting strongly with a two-level system in 
a dispersive regime without invoking the rotating-wave approximation (RWA). First we recast the adiabatic approximation 
by applying two successive unitary transformations. We also pointed out the frame toward the generalized rotating-wave  approximation (GRWA) that improves the adiabatic approximation~\cite{Irish07}.

Collapses and revivals in the two-level system are accompanied by beatings that are exhibited by the dynamics of the quantum oscillator. 
The beatings are related to collapses and revivals of the wave function in two-level system. 
However, on the much longer time scale the quantum oscillator shows true collapses and revivals that occur in the dynamics 
of the oscillator coordinate. Between the new collapses and revivals there are fractional revivals whose signature is present in both the  two-level 
system and the quantum oscillator. 

Finally, we calculated the scattered oscillator field off the two-level system. Due to the dispersive regime, the scattered field spectrum 
shows a doublet symmetrically displaced about the frequency of the unperturbed oscillator by the amount of the low frequency components of 
Rabi frequencies. The spectrum can be used not only to assess those Rabi frequencies but also to probe the distribution of the 
oscillator quantum number.  

\section{Acknowledgments}
The work has been supported by the Romanian Ministry of Education and Research under the project ``Ideas'' No.120/2007.




\end{document}